# Collisional Excavation of Asteroid (596) Scheila

D. Bodewits[1], M. S. Kelley[1], J.-Y. Li[1], W. B. Landsman[2], S. Besse[1], and M. F. A'Hearn[1]

[1] Department of Astronomy, U. Maryland, College Park, MD 20742. dennis@astro.umd.edu; msk@astro.umd.edu; jyli@astro.umd.edu; sbesse@astro.umd.edu; ma@astro.umd.edu

[2] NASA GSFC, Adnet Systems, Mailstop 667, Greenbelt, MD 20771. Wayne.B.Landsman@nasa.gov

4 figures, 1 table

**Abstract**
We observed asteroid (596) Scheila and its ejecta cloud using the Swift UV-optical telescope. We obtained photometry of the nucleus and the ejecta, and for the first time measured the asteroid's reflection spectrum between 290 – 500 nm. Our measurements indicate significant reddening at UV wavelengths (13% per $10^3$ Å) and a possible broad, unidentified absorption feature around 380 nm. Our measurements indicate that the outburst has not permanently increased the asteroid's brightness. We did not detect any of the gases that are typically associated with either hypervolatile activity thought responsible for cometary outbursts ($CO^+$, $CO_2^+$), or for any volatiles excavated with the dust (OH, NH, CN, $C_2$, $C_3$). We estimate that $6 \times 10^8$ kg of dust was released with a high ejection velocity of 57 m/s (assuming 1 μm sized particles). While the asteroid is red in color and the ejecta have the same color as the Sun, we suggest that the dust does not contain any ice. Based on our observations, we conclude that (596) Scheila was most likely impacted by another main belt asteroid less than 100 meters in diameter.

## 1. Introduction

Early December 2010, an unexpected dust cloud was discovered around the asteroid (596) Scheila (Larson et al. 2010). We report on observations using the UV-Optical Telescope (UVOT) on board Swift.

Asteroids have long been believed to be currently geologically inactive. Instead, their shapes, sizes, and surface geology are dominated by impacts. However, several asteroids have been observed to develop a coma and or dust tail (Hsieh & Jewitt 2006). Some asteroids are repeatedly active over periods as long as months, most famously 133P/Elst-Pizzaro (Hsieh et al. 2004). Others are only active in short-lived bursts. Two main ejecta processes have been suggested: cometary-like outgassing of volatiles driven by an exoergic interior process (Jewitt 2009), and inter-asteroidal collisions (Jewitt et al. 2010; Snodgrass et al. 2010). Alternatively, collisions may expose icy content buried beneath the surface (Hsieh et al. 2004).

(596) Scheila is a T-type main belt asteroid with a diameter of 113 km[1]. On 2010 Dec. 11.44 UT the asteroid had increased by $m_v$ = 0.8 and archival Catalina Sky Survey observations showed that the activity was triggered between 2010 Nov. 11 and Dec. 3 (Larson et al. 2010). The asteroid orbits the Sun in 5.0 yr with an eccentricity of 0.16 and a semi-major axis of 2.93 AU. Its outburst occurred at $r_h$ = 3.1 AU, close to its aphelion at 3.4 AU. Scheila offered a unique target as we observed it only weeks after its outburst; the outburst of P/2010 A2 (Linear) was discovered in January 2010 (Birtwhistle et al. 2010), almost a year after it was triggered (Jewitt et al. 2010; Moreno et al. 2010, Snodgrass et al. 2010).

## 2. Observations

Swift is a multi-wavelength observatory equipped for rapid follow-up of gamma-ray bursts (Gehrels et al. 2004). Our observations use the UVOT that provides a 17 × 17 arcminute field of view with a plate scale of 1 arcsec/pixel and a point spread function of 2.5" FWHM (Mason et al. 2004). Seven broadband filters allow color discrimination, and two grisms provide low-resolution ($\lambda/\delta\lambda$ = 100) spectroscopy between 170–650 nm.

Swift observed (596) Scheila on 2010 December 14 and 15 UT (Table 1). On both days we used the V ($\lambda_c$ 546.8 nm, FWHM 75.0 nm) and UVW1 ($\lambda_c$ 260.0 nm, FWHM 70.0 nm) filters. On Dec 15th we used the UV grism to search for gaseous emission lines. The grism was operated in 'clocked mode' to suppress background stars and the dispersion axis was oriented at a position angle of ~260°. The asteroid was not tracked, and the proper motion is 6.2" over our ~1300 s exposures.

The results are subject to several possible systematic uncertainties. The absolute calibration of UVOT[2] is accurate to within 0.01 magnitudes (V) and 0.03 magnitudes (UVW1). The brightness of the nucleus results in significant coincidence loss which we estimate to be 0.3 ± 0.05 mag at V and 0.02 ± 0.005 mag at UVW1. Shot noise and background subtraction uncertainty, as well as the aperture and coincidence loss correction errors are all included in the listed uncertainties.

---

[1] http://ssd.jpl.nasa.gov/sbdb.cgi
[2] http://swift.gsfc.nasa.gov/docs/swift/analysis/uvot_digest/zeropts.html

## 3. Results

### 3.1 Imaging
The images obtained with the V and UVW1 broadband filters were registered, binned, and logarithmically stretched to allow inspection of the morphology of the ejecta (Fig. 1). Both panels span about 215" x 204", corresponding to 390,000 x 370,000 km. We used archival Digitized Sky Survey[3] images to identify and remove background sources. The projected directions to the Sun and the orbital motion almost coincide and are indicated by arrows.

We have clearly resolved detections of the ejecta in both filters. In the V-band, the morphology seems to consist of a broad 'northern' fan from Position Angle (PA) = 330 to 60°, and a curved 'southern' feature at PA = 180°.

Aperture photometry on the images yields an average $m_v$ = 14.1 ± 0.07 and $m_{uv}$ = 16.4 ± 0.09 for the asteroid in a 10 pixel radius aperture. We made no attempt to remove the dust emission within our asteroid aperture. Subtracting the nucleus' brightness, the total dust brightness in a 160" × 100" aperture is $m_v$ = 14.4 ± 0.08. The northern plume is brighter than the southern plume (65 vs. 35% of the total ejecta flux). The ejecta are faint in UVW1 (Fig. 1) and best measured in a smaller 80" x 80" aperture to minimize uncertainties from the background subtraction. We find an average $m_v$ = 14.8 ± 0.08 and $m_{uv}$ = 16.8 ± 0.11 for the ejecta in the smaller aperture (Table 1 and Figure 1). Our measurements are close to the asteroid's predicted brightness ($m_v$ = 14.21[4], with an estimated maximum amplitude of 0.09[5]). Within the uncertainties, the ejecta (UVW1–V = 1.91 ± 0.1) have solar colors (UVW1–V = 1.94), while the asteroid (UVW1–V = 2.36 ± 0.11) is redder than the Sun.

### 3.2 Spectroscopy
Figure 2 shows the asteroid's spectrum extracted from a rectangular region 13 pixels wide along with a solar spectrum obtained on 2010 Dec. 15 (SORCE; McClintock et al. 2005[6]). The spectrum below 280 nm was contaminated by a background star. The data were binned by a factor of 8 to improve the signal to noise, and to suppress the Mod8 pattern in the grism image. For comparison, the grism spectrum of comet C/2007 N3 (Lulin) is shown in Fig. 2 (Bodewits et al. 2011). While Lulin's spectrum is dominated by the emission features of several gaseous species (e.g. OH, CN, $C_2$, $C_3$, NH, $CO_2^+$), there is no evidence of any gaseous emission feature in the asteroid's spectrum. Assuming an OH fluorescence feature similar to that in Lulin's spectrum (50 nm wide due to the gas distribution), a 1σ flux upper limit of $1 \times 10^{-13}$ erg s$^{-1}$ cm$^{-2}$ A$^{-1}$, and a total OH emission fluorescence of $2.2 \times 10^{-16}$ erg s$^{-1}$ cm$^{-2}$ molecule$^{-1}$ (Schleicher & A'Hearn 1988) yields a an upper limit $3 \times 10^{30}$ OH molecules around the nucleus. Considering the long lifetime of OH at 3.1 AU from the Sun ($\sim 10^6$ s, Huebner et al. 1992), the water production rate would be <$10^{25}$ molecules s$^{-1}$, 1000 times lower than typically observed in Jupiter family comets at 1 AU. Although we cannot constrain the gaseous output at the onset of the anomaly, we conclude that the asteroid did not produce significant OH or water gas during the observations.

---

[3] http://archive.stsci.edu/cgi-bin/dss_form
[4] http://ssd.jpl.nasa.gov/?horizons
[5] http://sbn.psi.edu/ferret/
[6] http://lasp.colorado.edu/lisird/sorce/sorce_ssi/index.html

Scaling the solar spectrum yields a best fit of m$_v$ = 14.3 ± 0.1, in agreement with our broadband observations. We derived the asteroid's albedo using Equation 2, assuming $\Phi(\alpha)$ = 0.45, a diameter of 113 km (Dunham & Herald, 2005[5]), and the solar irradiance from SORCE (Fig 3). We combined this with a scaled reflectance spectrum (Bus & Binzel 2002[4]). Our data are in excellent agreement with albedo measurements from other observations (Dunham & Herald, 2005[5]; Ryan & Woodward 2010). Scheila's spectrum is significantly reddened compared to the Sun.

There appears to be a broad absorption feature between 320 – 420 nm, centered at about 380 nm, superimposed on the red slope. It is difficult to interpret the UV spectrum of Scheila due to both the limited number of asteroids observed at this wavelength (Butterworth and Meadows, 1985; Roettger and Buratti, 1994), and due to the lack of laboratory reflectance spectra of possible compositional minerals. UV absorptions at slightly shorter wavelengths are present with various intensities for possibly hydrated asteroids such as (1) Ceres, (2) Pallas, and (102) Hygiea (Butterworth and Meadows, 1985; Roettger and Buratti, 1994; Li et al., 2006; 2009; Rivkin et al., 2006; Milliken and Rivkin, 2009). UV spectra of S-types show a strong red slope and sharp drop at 300 to 400 nm, and very weak UV absorptions. As a primitive type asteroid in the outer main asteroid belt with ~4% albedo, Scheila probably has a surface composition that is closer to carbonaceous meteorites. In order to better understand the UV spectra of asteroids, observations of various spectral types and laboratory measurements of carbonaceous meteorites are needed.

## 4. Discussion

### 4.1 Ejecta Dynamics

The sunward extend of the cloud is ~15", corresponding to 2.6 × 10$^4$ km at the asteroid. This is the distance X$_R$ where solar radiation turns dust particles, and it can be used to estimate the grain ejection velocity (Hsieh et al 2004, Jewitt & Meech 1987):

$$X_R \sim \frac{v_d^2 r_h^2}{2\beta_d g_{sun}} \qquad 1$$

where v$_d$ is the relative ejection velocity, r$_h$ the heliocentric distance in AU, β the dimensionless ratio between radiation pressure acceleration and the local gravity, and g$_{sun}$ = 0.006 m s$^{-2}$ is the gravitational acceleration to the Sun at 1 AU. Using this we find v$_d$ = 57$\sqrt{\beta}$ m s$^{-1}$.

This velocity is an order of magnitude larger than those reported for 133P (1.5$\sqrt{\beta}$ m s$^{-1}$; Hsieh et al. 2004) and P/2010 A2 (1.1$\sqrt{\beta}$ m s$^{-1}$ Moreno et al. 2010). This difference is partially explained by the different escape velocities. Assuming a density of 2500 kg m$^{-3}$, and the estimated diameters (113 km vs. 0.12 km), the escape velocities are 75 m s$^{-1}$ for 596 and 0.06 m s$^{-1}$ for P/2010 A2. Additionally, the collision disrupting A2 may have produced a lot of high velocity dust, but this had left the vicinity of A2 over the year between the impact and its discovery.

We simulated the ejection of dust from 596 to better understand the dynamics of the observed ejecta. The simulation accounts for the gravitational acceleration of the Sun and planets and solar radiation pressure (Kelley 2006; Kelley et al. 2009). The model was designed to simulate observations of comet comae, and does not account for the mass of the parent body. Since 596 is a large asteroid, our simulation is only a rough approximation to the true dust ejection dynamics. We ejected $10^6$ grains isotropically from the surface in a single event, with a distribution of sizes (0.1 – $10^4$ μm) and velocities (0 – 100 m s$^{-1}$) on 2010 Dec. 3, and integrated their positions forward in time to 2010 Dec. 15. The full extent of the dust in our UVOT images can be encompassed in a circle centered on the asteroid with a radius of 76" (140,000 km). Our simulations show that 12 days after the outburst, only grains >1 μm in radius remain within this aperture. Therefore, we adopt 1 μm as a lower-limit to the ejected grain size. A stricter limit to the grain radii can be estimated if the outburst date can be better constrained, and if the dynamics of the ejecta could be reproduced, neither of which have we attempted. Independent of these refinements, all grains smaller than ~1 μm would be found outside the observed extent of the dust in our V-band image. Therefore, we conclude that sub-micron sized grains are few or absent in the ejecta.

We continued our simulation forward in time with 10-day steps to estimate the lifetime of the ejecta within our 76" radius aperture. We have weighted the dust grain cross sections by v$^{-1}$ to approximate crater formation dynamics (more mass is ejected at lower velocities), and converted cross section to V-band magnitudes using Eq. 2 (Figure 4). The evolution of the dust results in a rapid decline of $m_v$ of approximately 3 magnitudes over the first month. If the ejecta are devoid of large grains (i.e., the grain size upper-limit is 10 μm, rather than our initial assumption of $10^4$ μm), then the dust will be even fainter, falling 5 magnitudes by mid-January. According to our simulations, unless the grain radii extend at least into the millimeter size range, we expect that the surface brightness of the ejecta cloud around Scheila will be too faint to observe in early 2011.

### *4.2 Dust Mass*

The total reflection cross section $C_d$ is related to the difference between the V-band magnitudes of dust and the apparent solar magnitude ($m_{sun}$ = -26.74) by the following relation:

$$C_d = 2.25 \times 10^{22} \pi \frac{r_h^2 \Delta^2 10^{-0.4(m_d - m_{sun})}}{p_V \Phi(\alpha)} \qquad 2$$

where the heliocentric distance $r_h$ = 3.1 AU, the geocentric distance $\Delta$ = 2.5 AU, $p_V$ is the geometric albedo, and $\Phi(\alpha)$ the phase darkening. Assuming a phase correction of 0.75 for a solar phase angle of 16 degrees (Schleicher et al. 1998), and a dust albedo of 0.1 (Jewitt 2009), we find $C_d$ ~ 2 × 10$^9$ m$^2$ for the ejecta. Further assuming a power law distribution of spherical particles with radii between $10^{-6}$ – $10^{-2}$ m

(section 4.1), with a slope of -3.5, and a density of 2500 kg m$^{-3}$, we can use the relation between mass and scattering cross section derived by Jewitt (2009). We then find that there was approximately 6 × 10$^8$ kg of dust around the asteroid.

To put this in perspective, this is a much larger mass than found around active centaurs (10$^4$ – 10$^6$ kg, Jewitt 2009), comparable to the mass released by the destruction of P/2010 A2 (LINEAR) (5 – 60 × 10$^7$ kg; Moreno et al. 2010, Snodgrass et al. 2010, Jewitt et al. 2010), and comparable to the total amount of dust and gas excavated by the Deep Impact probe colliding with 9P/Tempel 1 (10$^6$ – 10$^9$ kg, Küppers et al. 2005; A'Hearn et al. 2008).

### *4.3 Asteroid Impact Scenario*

We did not detect any evidence of volatiles that are typically associated with the hypervolatile activity (CO$^+$, CO$_2^+$), or of any volatiles that were possibly released from the ejected dust (OH, NH, CN, C$_2$, C$_3$). The reddening of the asteroid (13% per 10$^3$ Å) and the neutral color of the ejecta are comparable to those of comet surfaces and comae, which are generally considered to be ice-free. Comets 9P/Tempel-1, 21P/Giacobini-Zinner, and 67P/Churyumov-Gerasimenko have nucleus spectral reflectivity gradients of 12.5 ± 1%, 12.8 ± 2.7%, and 11 ± 2%, and comae gradients of about 8%, 8 ± 3%, and -1 to 1%, respectively (Li et al. 2007; Luu et al. 1993; Tubiana et al. 2008; Schleicher et al. 2007; Kolokolova et al. 1997; Weiler et al. 2004; Lara et al. 2005). We therefore conclude that there were no detectable amounts of gas or ice around the asteroid during our observations. However, most fragment species observed in optical/UV wavelengths have lifetimes corresponding to about 11.1 days at the heliocentric distance of the asteroid (Huebner et al. 1992). Icy grains have even shorter lifetimes (Bockelée-Morvan et al. 2001). While we cannot rule out that gas or ice was released at the onset of the outburst (before Dec 3$^{rd}$), our results do imply that the excavation event did not trigger any continuous activity.

The plumes are consistent with ejecta moving away from the asteroid and then being pushed along the anti-solar direction by solar radiation pressure. The observed morphology could be well explained by an oblique impact where one plume would be associated with downrange ejecta and the other with material ejected vertically due to the mechanical excavation by shockwaves generated by the impact (see e.g. Schultz et al. 2007).

We estimated above that about 6 × 10$^8$ kg of dust was released. Assuming a density of 2500 kg m$^{-3}$, this corresponds to a sphere with a diameter of 80 meters. If we assume that most of the momentum input from the impact is carried out by the ejecta, and an impact velocity of 5 km s$^{-1}$, we find an impactor diameter of 20 m, consistent with our first estimate. If the observed dust was indeed ejected by an impact, then the ejecta probably constitute only 10 – 20% of the total excavated mass. The mass of the material excavated by hypervelocity impacts usually exceeds the mass of the projectile, depending on many parameters such as the impact velocity and geometry, porosity, and the strength of the target (Holsapple et al. 2002, Schultz et al. 2005). We conclude that the projectile was therefore probably <100 m in diameter.

How likely is this collision scenario? Asteroid collisions have been extensively studied for evolution studies (Bottke et al. 1994, Davis et al. 2002). In the main belt, the most likely collision partners are other main belt asteroids. Bottke et al (1994) give an intrinsic collision probability of $3 \times 10^{-18}$ yr$^{-1}$ km$^{-2}$ at a relative velocity of 5.3 km s$^{-1}$. Multiplying this with the square of the radius of Scheila and N, the number of projectiles sized 10 – 100 meter yields the impact frequency. A compilation of different distribution models suggests that the number of projectiles is in the range $10^{10} – 10^{11}$ (Davis et al. 2002). Based on this estimate, an object the size of (596) Scheila would be impacted by a small asteroid once every 1,000 years. There are ~200 asteroids the size of (596) Scheila (Tedesco et al. 2004), thus an event like this would occur once every 5 years, and collisions with asteroids smaller than 10 meter should occur even more often. Based on statistics, a collisional impact is thus very likely.

More tentatively, it is possible to estimate what effect the impact had on the surface of Scheila. Craters on (243) Ida, (951) Gaspra, (25143) Itokawa, (2867) Steins, and (443) Eros have different depth-to-diameter ratios that are linked to the history of the asteroid itself (Sullivan et al 1996, Carr et al. 1994, Hirata et al. 2009, Besse et al. 2011, Robinson et al. 2002). Assuming spherical craters, the diameter D is related to the volume $V \sim 0.06\ D^3$ (Scheeres et al. 2002; depth-to-diameter ratio of 0.15). Based on the equivalent volume of the ejecta ($3 \times 10^5$ m$^3$), and considering that the ejected mass may only constitute 10 – 20% of the total excavated mass, the impact may have resulted in a crater with a diameter of 300 meters. Fresh craters often have a higher depth-to-diameter ratio than aged craters, consequently the impact may well have created a crater with a diameter much larger than 300 meters. An impact of that size is not unexpected on the surface of asteroids. During its recent flyby of the asteroid (21) Lutetia, which has a size comparable to (596) Scheila, the Rosetta spacecraft discovered numerous craters with diameters larger than 500 meters, likely the scars of impacts similar to the one that occurred on (596) Scheila.

## 5. Conclusions

We obtained optical and UV images and spectroscopy of asteroid (596) Scheila using the Swift UVOT within at most 6 weeks after it first started to show a dramatic increase in brightness. We obtained the first measurement of its reflectance spectra below 450 nm, a wavelength regime in which asteroids are not very well explored. Whatever caused the ejection of dust from Scheila has not permanently increased the asteroid's brightness. We did not detect any of the gases that are typically associated with either hypervolatile activity thought responsible for cometary outbursts (CO$^+$, CO$_2^+$), or for any volatiles excavated with the dust (OH, NH, CN, C$_2$, C$_3$). We estimate that $6 \times 10^8$ kg of dust was released with an ejection velocity of 57 m s$^{-1}$ (assuming 1 micron sized particles). The ejecta have the same color as the Sun, while the asteroid appears reddened. The two colors are consistent with observations of comet comae and nuclei, both generally considered to be ice-free. Based on our observations, we conclude that Scheila was most likely impacted by another main belt asteroid with a diameter of at most 100 m.

## Acknowledgements
We thank the Swift team for use of Director's Discretionary Time and for the careful and successful planning of these observations. We thank Pete Schultz for valuable discussions on this study.
## References

A'Hearn, M. F. 2008. Space Sci. Rev. 138, 237

Besse, S., et al. 2011, Icarus submitted

Birtwhistle, P., Ryan, W. H., Sato, H., Beshore, E. C., & Kadota, K. 2010, IAUC, 9105, 1

Bodewits, D., Villanueva, G.L., Mumma, M.J., Landsman, W.B., Carter, J.A., and Read, A.M. 2011, AJ, 141, 12.

Bockelée-Morvan et al. 2001, Science 292, 1339

Bottke, W.F., Nolan, M.C., Greenberg, R., Kolvoord, R.A. 1994, Icarus, 107, 255

Bus, S.J., and Binzel, R.P., 2002, Icarus 158, 1

Butterworth, P.S., and Meadows, A.J., 1985, Icarus 62, 305-318

Carr, M.H., et al. 1994, Icarus 107, 61

Davis, D.R., Durda, D.D., Marzari, F., Campo Bagatin, and Gil-Hutton, R. 2002, in Asteroids III. Ed. Bottke Jr., W.F., Cellino, A., Paolichhi, P., and Binzel, R.P. (Tuscon, AZ: Univ. Arizona Press), 545

Dunham, D. W. and Herald, D., 2005, Asteroid Occultation List. EAR-A-3-RDR-OCCULTATIONS-V3.0:OCCLIST_TAB. NASA Planetary Data System

Gehrels, N., et al. 2004, ApJ, 611, 1005

Hirata, N. et al. 2009, Icarus 200, 486-502

Holsapple, K., Giblin, I., Housen, K., Nakamura, A., and Ryan, E. 2002, in Asteroids III. Ed. Bottke Jr., W.F., Cellino, A., Paolichhi, P., and Binzel, R.P. (Tuscon, AZ: Univ. Arizona Press), 443

Hsieh, H.H., & Jewitt, D.C. 2006, Science, 312, 561

Hsieh, H.H., Jewitt, D.C., and Fernandez, Y.R. 2004, ApJ, 127, 2997

Huebner, W.F., Keady, J.J., and Lyon, S.P. 1992, Astroph. Sp. Sci., 195, 1

Jewitt, D.C., & Meech, K.J. 1987, ApJ 317, 992

Jewitt, D.C. 2009, AJ, 137, 4296.

Jewitt, D., Waever, H.A., Agarwal, J., Mutchker, M., Drahus, M., Nature (2010) vol. 467 (7317) pp. 817-819

Kelley, M. S. 2006, Ph.D. Thesis, University of Minnesota, Minneapolis, MN

Kelley, M. S., Wooden, D. H., Tubiana, C., Boehnhardt, H., Woodward, C. E., & Harker, D. E. 2009, AJ, 137, 4633

Kolokolova, L., Jockers, K., Chernova, G., Kiselev, N., 1997, Icarus 126, 351.

Küppers, M., et al. 2005, Nature, 437, 987.

**Figures & Tables**

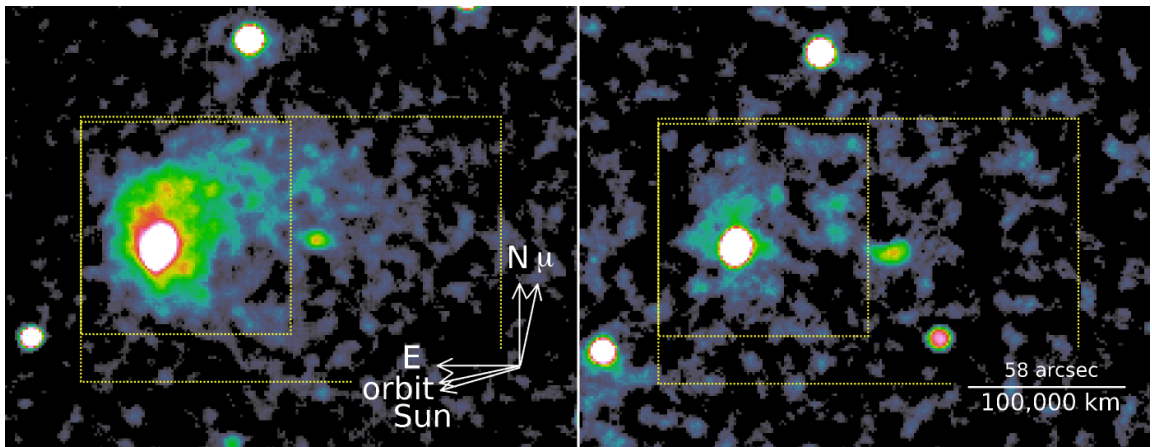

**Figure 1.** UVOT images obtained with the V (left) and UVW1 (right) filters on Dec 15th 2010 UT. The physical scale and orientation is the same in both frames. Both images are scaled logarithmically, but stretched and smoothed separately to enhance the ejecta. The untracked asteroid appears slightly elongated in the northward direction (μ). Two plumes can be seen, and their morphology is consistent with ejecta moving away from the asteroid and then being pushed in the anti-solar direction by solar radiation pressure. The yellow boxes indicate the area from which we obtained the color (small) and total (large) photometry.

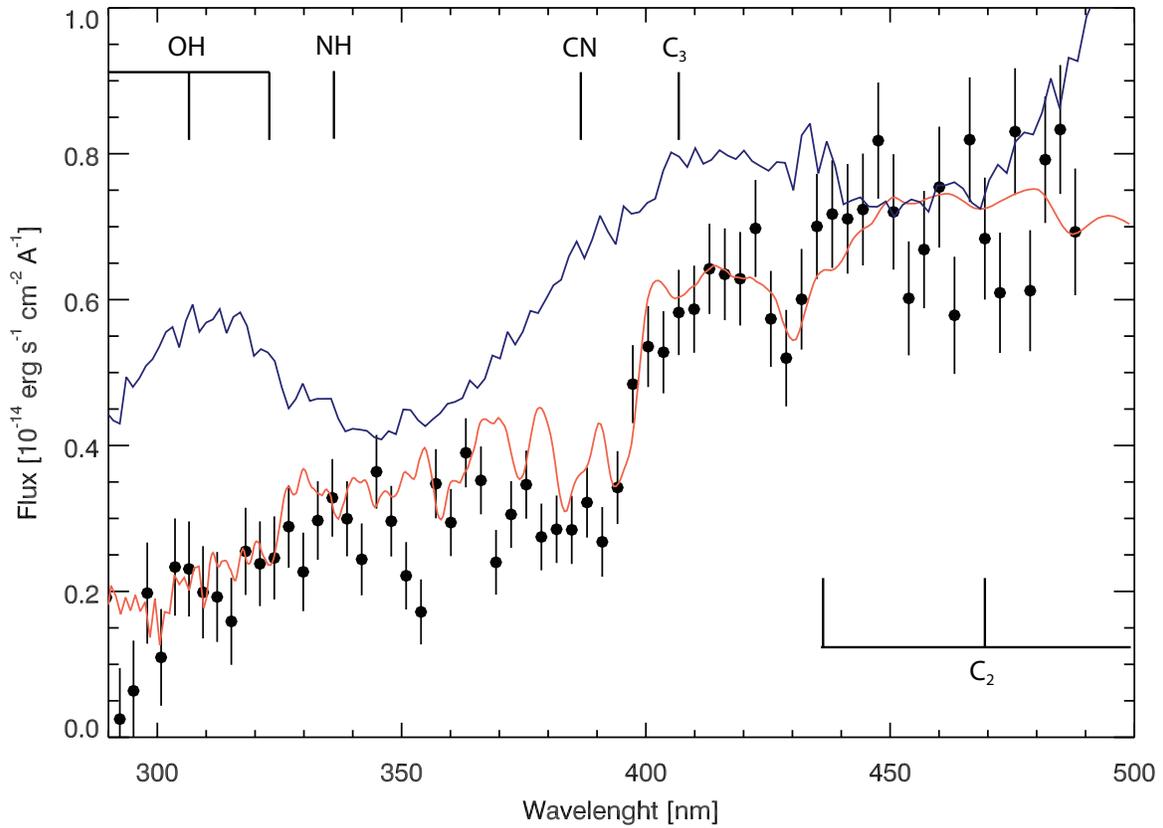

**Figure 2.** Spectra of Scheila (black), comet C/2007 N3 (Lulin) (blue; Bodewits et al. 2011) normalized at 450 nm) and the Sun (red; McClintock et al. 2005), scaled to $m_v$ = 14.3. Scheila's spectrum is reddened compared to the Sun. There is no evidence of volatile emission lines that dominate Lulin's spectrum (labeled). Error bars indicate 1σ stochastical errors; the systematic uncertainty is approximately 25%.

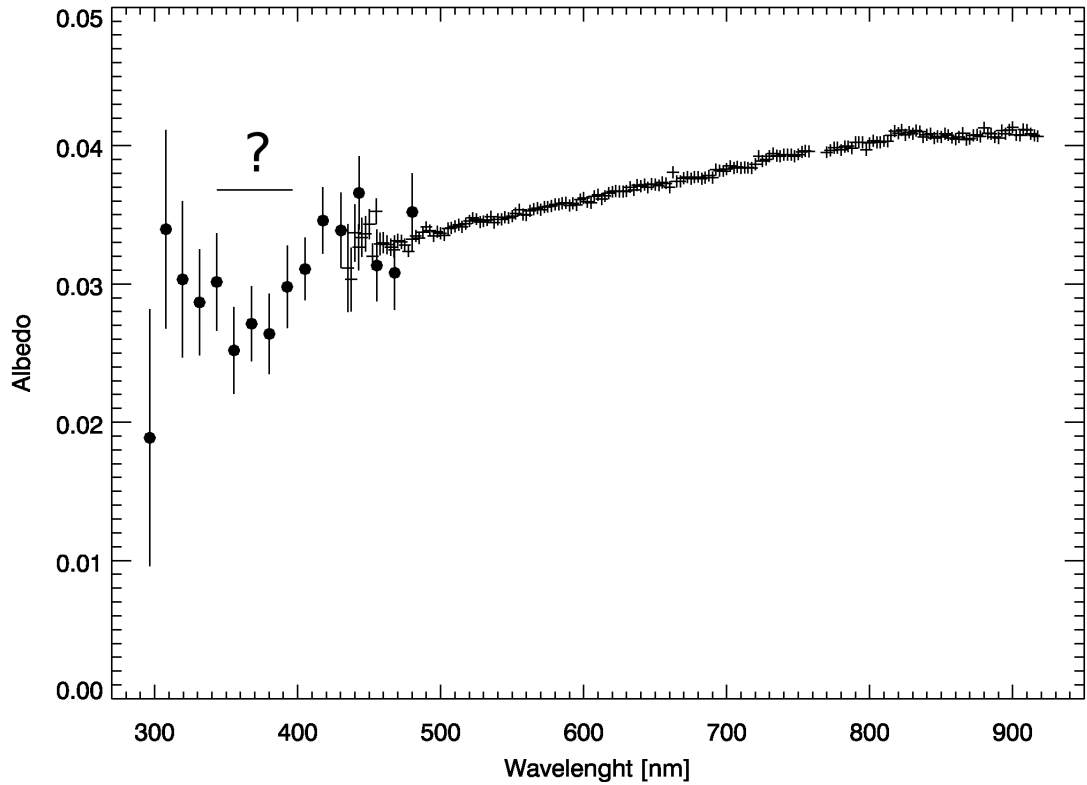

**Figure 3.** The albedo of Scheila. (•) Swift measurements; (+) Reflectance measurements by Bus & Binzel (2002), scaled to match our albedo measurement. Error bars indicate 1σ stochastical errors. Our data are in excellent agreement with other albedo measurements. We tentatively identify a broad absorption feature between 320–420 nm, indicated by the question mark.

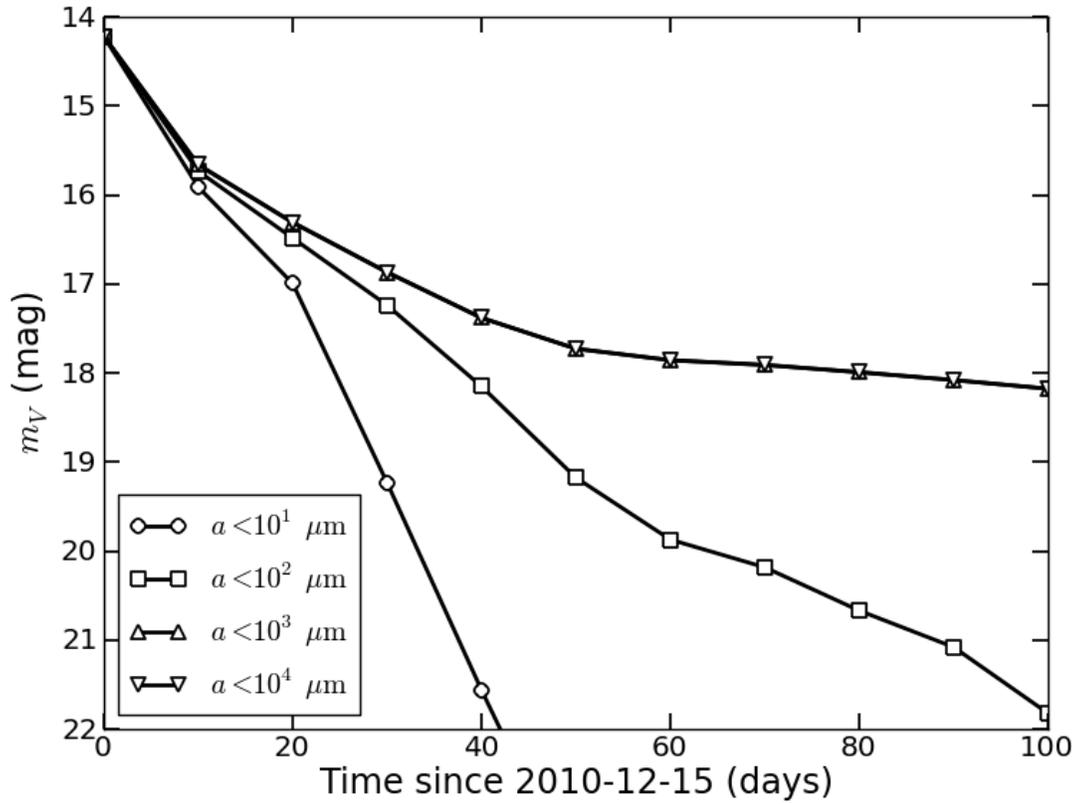

**Figure 4:** Simulated apparent V-band magnitude of dust within 76". Light curves are given for grain radius upper-limits of 10 to $10^4$ μm. The larger radiation pressure efficiency of smaller grains removes them from the field of view quickly, reducing the reflection cross section of the dust. We expect that the surface brightness of the ejecta cloud will soon be too faint to observe.

**Table 1**: Observing log and photometry results.

| Start Observing Time (UT) | Filter | Exp. Time (seconds) | m Nucleus[a] | m Ejecta[a,b] (160"×100") | m Ejecta[a,b] (80"×80") |
|---|---|---|---|---|---|
| 2010-12-14T06:47 | V | 1285 | 14.1 ± 0.05 | 14.4 ± 0.06 | 14.8 ± 0.06 |
| 2010-12-14T07:09 | UVW1 | 701 | 16.4 ± 0.07 | - | 16.7 ± 0.08 |
| 2010-12-15T06:56 | UV grism | 1072 | 14.3 ± 0.1 | - | - |
| 2010-12-15T08:29 | V | 1300 | 14.0 ± 0.05 | 14.3 ± 0.06 | 14.9 ± 0.06 |
| 2010-12-15T08:50 | UVW1 | 1087 | 16.3 ± 0.04 | - | 16.8 ± 0.08 |

**Notes:**
[a] Errors include both stochastic and absolute uncertainties.
[b] The UVW1 ejecta is faint and best measured in the smaller 80" x 80" aperture to minimize background stars and the systematic uncertainty.